\documentclass[sigplan,nonacm]{acmart}
\acmSubmissionID{<509>}
\renewcommand\footnotetextcopyrightpermission[1]{}
\usepackage{times}  
\usepackage{hyperref}
\usepackage{tikz}
\usepackage{amsmath}
\usepackage{graphicx}
\usepackage{xspace}
\usepackage{xcolor}
\usepackage[T1]{fontenc}
\usepackage{subfigure}
\usepackage{verbatim}
\usepackage{url}
\usepackage{enumitem}
\usepackage{listings}
\usepackage{multirow}
\usepackage[ruled,linesnumbered]{algorithm2e}
\usepackage{pifont}
\usepackage{mathtools}
\usepackage{fancyhdr}

\newcommand{\sysname}{FFTrainer\xspace}
\newcommand{\ccl}{LCCL\xspace}

\newcommand{\para}[1]{\noindent\textbf{#1}}
\newcommand{\code}[1]{{\small{\texttt{#1}}}\xspace}
\newcommand{\num}[1]{\normalsize{\textcircled{\scriptsize{#1}}}\normalsize\xspace}

\newcommand{\ckpt}{{\small{\texttt{CKPT}}}\xspace}
\newcommand{\traintransfer}{{\texttt{TRAIN}}\xspace}
\newcommand{\statetransfer}{{\texttt{STATE}}\xspace}

\newcommand{\tid}{{\texttt{TID}}\xspace}

\AtBeginDocument{%
  }

\setcopyright{acmlicensed}
\copyrightyear{2025}
\acmYear{2025}
\acmDOI{XXXXXXX.XXXXXXX}

\acmConference[SOSP'25]{ACM Conference}{Oct 2025}{Seoul, Korea}%
\acmISBN{978-1-4503-XXXX-X/18/06}

\begin{document}

\title{\sysname: Fast Failover in Large-Language Model Training with Almost-Free State Management}


\author{Bohan Zhao$^\dagger$}
\affiliation{
  \institution{Tsinghua University}
  \country{}}
\email{zhaobh23@mails.tsinghua.edu.cn}

\author{Yuanhong Wang$^\dagger$}
\affiliation{
  \institution{Tsinghua University}
  \country{}}
\email{yuanhong22@mails.tsinghua.edu.cn}

\author{Chenglin Liu}
\affiliation{
  \institution{Tsinghua University}
  \country{}}
\email{liucl22@mails.tsinghua.edu.cn}

\author{Jiaqi Pan}
\affiliation{
  \institution{Tsinghua University}
  \country{}}
\email{panjq22@mails.tsinghua.edu.cn}

\author{Guang Yang}
\affiliation{
  \institution{Tsinghua University}
  \country{}}
\email{yangg22@mails.tsinghua.edu.cn}

\author{Ruitao Liu}
\affiliation{
  \institution{Tsinghua University}
  \country{}}
\email{liurt23@mails.tsinghua.edu.cn}

\author{Tingrui Zhang}
\affiliation{
  \institution{Tsinghua University}
  \country{}}
\email{zhangtr22@mails.tsinghua.edu.cn}

\author{Kai Luo}
\affiliation{
  \institution{Tsinghua University}
  \country{}}
\email{luok22@mails.tsinghua.edu.cn}

\author{Wei Xu\textsuperscript{*}}
\affiliation{
  \institution{Tsinghua University}
  \country{}}
\email{Weixu@tsinghua.edu.cn}

\thanks{$^\dagger$Equal contribution.}
\thanks{\textsuperscript{*}Corresponding author.}





\renewcommand{\shortauthors}{}


\begin{abstract}
Recent developments in large language models (LLMs) have introduced new requirements for efficient and robust training.  As LLM clusters scale, node failures, lengthy recoveries, and bulky checkpoints erode efficiency. Infrequent asynchronous checkpoints trigger costly rollbacks, yet higher frequencies add prohibitive overhead.  To address these challenges, we propose \sysname, a system designed for robust LLM training. \sysname leverages surplus network capacity to quickly save and load states, thereby preventing rollbacks and accelerating recovery. Compared with prior checkpointing approaches, \sysname reduces recovery time by up to 98\% and mitigates GPU utilization loss by up to 68\% without hindering normal training.  
\end{abstract}


\maketitle

\section{Introduction}
\label{sec:intro}

Large language models (LLMs) are widely used for tasks like question answering, content generation and coding assistance~\cite{chatgpt,Copilot}, drawing significant attention from both academia and industry.  Training these models requires large computation clusters involving substantial investments.  

Unfortunately, these expensive clusters face frequent and lengthy downtime~\cite{jiang2024megascale}, leading to significant time and financial losses. For instance, a 16,384-GPU cluster used for training LLaMA3.1 405B reports a \emph{mean time between failure (MTBF)} of about three hours. Even worse, \emph{mean time to recover (MTTR)} from such failures can be tens of minutes~\cite{jiang2024megascale}, halting training progress. Studies show that failures can slow down training progress by up to 43\%~\cite{maeng2021partial}.

The training process design directly causes the short MTBF. Distributed LLM training relies on tightly coupled \emph{workers} that communicate using MPI-style collective operations, which are not failure-tolerant. Thus, when any worker fails, the entire job must restart from a \emph{checkpoint} (\ckpt).  Even with GPUs that have an MTBF of around 80,000 hours (9 years), the MTBF for a cluster of 16,000 GPUs reduces to just three hours.  While we can try to support Spark~\cite{zaharia2010spark}-style partial recovery,  managing the partial states would complicate the already-complex training process and add overhead during normal operation.

A more practical and economical approach is to tolerate the short MTBF by reducing the MTTR to seconds while introducing minimal overhead during normal execution.
To do so, we first perform a comprehensive study to identify the main causes of the long MTTR.  Most recovery time is spent recovering essential \emph{states} on workers. These states include both training states like input data, model weights, and optimizer states, and network states. Unfortunately, in current training frameworks, neither kind recovers quickly.  

The main challenge of recovery from a \ckpt is the large \emph{size} of the states. In state-of-the-art training libraries like DeepSpeed~\cite{DeepSpeed}, each GPU needs to checkpoint tens of GBs of states. Loading them from/to storage not only delays recovery but also consumes significant disk, PCIe, and network bandwidth, further slowing down training.
\emph{Multi-level asynchronous checkpointing} is widely used in recent solutions like Gemini~\cite{wang2023gemini} and ByteCheckpoint~\cite{wan2024bytecheckpoint}, where states are first saved to main memory in a pipeline fashion before being written to remote disks by background threads.  However, due to network and disk bandwidth limitations, completing a single {\ckpt} can take several iterations. The background thread keeps consuming network and I/O resources, causing non-negligible slowdowns (see Section~\ref{sec:theory}). 
These approaches do not fully exploit the redundancy in LLM {\ckpt}, resulting in unnecessary overhead for saving training states. 

In addition to the overhead that constrains {\ckpt} frequency, state-of-the-art systems miss key recovery optimizations.
Recovering network connection states involves multi-round communications to initialize communication groups and establish connections at various layers, such as PyTorch~\cite{pytorch} agents and NCCL~\cite{nccl} \emph{ranks}, with the process requiring over 10 minutes on large clusters~\cite{jiang2024megascale}. 
Additionally, there is an unnecessary dependency between network and training state recovery since worker IDs in the training framework reuse network ranks, causing delays when PyTorch assigns worker IDs after communication initialization, which can take over 1,000 seconds with 2,048 GPUs, thus delaying the recovery process.
Moreover, training rarely saturates the network. Modern cluster designs provision ample fabric to absorb synchronous bursts (e.g., \emph{all-to-all}, \emph{allreduce})\cite{Hyperplane}, but links are largely idle otherwise. On an $8{\times}4090$ server with a single $200 \text{Gbps}$ NIC, across four models (Table\ref{tab:data}), the average per-iteration utilization is only 1–3\%.

\begin{table}[t]
\caption{Data input and output per iteration of eight RTX 4090 when training with a 200 Gbps NIC.}
\label{tab:data}
\resizebox{\columnwidth}{!}{%
\begin{tabular}{|c|c|c|c|c|}
\hline
\textbf{Model}                     & \textbf{GPT-2} & \textbf{LLaMA3-8B} & \textbf{LLaMA2-13B} & \textbf{LLaMA3-70B} \\ \hline
\textbf{Per-iter time (s)}             & 21            & 11               & 36                 & 77                  \\ \hline
\textbf{NIC capacity (GB)}    & 525           & 275               & 900                 & 1925                \\ \hline
\textbf{Data input (KB)} & 128             & 64               & 64                 & 32                \\ \hline
\textbf{Data output (GB)} & 12            & 9.1                & 13                 & 35                 \\ \hline
\end{tabular}%
}
\vspace{-5pt}
\end{table}

Based on these observations, we present \sysname that eliminates the unnecessary bottlenecks in failover, reducing the MTTR to \emph{tens of seconds}, and utilizing otherwise unused network bandwidth for {\ckpt} and training data distribution with almost zero performance impact on normal training.  
Specifically, key ideas of \sysname include:

(1) \textbf{Only {\ckpt} necessary information:} Many states are naturally redundant across workers (e.g., within the same data parallel group). Unlike existing solutions, \sysname only backs up unique states, enabling \emph{instant checkpointing} (i.e., {\ckpt} at \emph{every} iteration).

(2) \textbf{Decouple training roles from network ranks:} 
\sysname decouples the \emph{logical role} of a worker in the training job (\code{role} $=(id1,id2,id3)$ in data/tensor/pipeline parallel groups) from its ranks in the NCCL layer.  Therefore, training state recovery becomes independent of network state recovery, allowing us to overlap them.

(3) \textbf{Reduce network states to recover:} We observe that many network states (e.g., topologies and communication groups) in NCCL are unnecessary for LLM, slowing down initialization.  Thus, we design a \emph{lightweight collective communication library (LCCL)} to simplify the stack and speed up cross-node reconnections.

(4) \textbf{Leveraging unused network bandwidth for the free transfer of {\ckpt}s and training data:}  With the software control of \ccl, \sysname manages worker traffic without hardware QoS support, transferring {\ckpt}s and training data only during communication idle periods, avoiding slowing down training.
This also eliminates the need for a dedicated data network or pre-partitioning data onto local storage.


(5) \textbf{Compatible and easy integration into existing frameworks:}
We integrated \sysname with the popular PyTorch framework and libraries like Megatron~\cite{narayanan2021megatron} and DeepSpeed. \sysname only introduces dozens of lines of code changes (mainly changing the package names for {\ckpt} and collective communication),  enabling compatibility with off-the-shelf training code on existing models.


Compared to existing solutions, \sysname asynchronously manages state transfer and reduces state sizes by exploiting redundancy in the training framework. Experiments with four LLMs of different sizes on a 128-GPU cluster (16 servers) with a commodity network show that \sysname eliminates {\ckpt} overhead, enables instant checkpointing, reduces job recovery time from about 1,000 seconds to 29 seconds, and removes the need for a dedicated storage network. Our analysis indicates these gains persist across a wide range of future compute and network infrastructures.

In summary, our contributions include:
(1) We diagnose design flaws and bottlenecks that slow checkpointing and recovery. 
(2) We introduce fast failure detection, near-zero-cost checkpointing, and efficient training data management that exploits otherwise unused network bandwidth.
(3) We integrate \sysname into PyTorch, Megatron, and DeepSpeed bundles, requiring minimal changes to training code.
(4) Real-cluster experiments show almost $10\times$ faster recovery time with negligible overhead on normal training. 
\emph{When the paper is published, we will open \sysname source code to the community.}

\section{Background and Related Work}
\label{sec:background}


\para{Distributed training (DT) and 3D parallelism.  }
DT is crucial for modern NLP tasks, which rely on large transformer-based models~\cite{vaswani2017transformer} like LLaMA~\cite{touvron2023llama, dubey2024llama3.1}, GPT-3~\cite{ouyang2022gpt,brown2020gpt3}, and BERT~\cite{devlin2018bert} for improved accuracy. DT is used to handle the growing size of models and data, with container orchestration systems like Kubernetes~\cite{Kubernetes} managing \emph{job} submission and software deployment on GPU servers. When a job is created, the scheduler selects physical nodes based on resource availability and constraints, assigning each to a \emph{pod} that runs the job's applications, libraries, and dependencies in a container. \emph{3D parallelism} combines both data and model parallelism. In data parallelism, each worker processes a batch of training data on a single GPU, calculating local \emph{gradients} at each iteration. Workers form a \emph{parallel group} and synchronize their models by aggregating gradients using \emph{parameter servers (PS)}~\cite{li2013parameter, li2014ps, jiang2020unified} or \textit{ring allreduce}~\cite{dean2012model, patarasuk2009allreduce}. Model parallelism, including \emph{pipeline parallelism} and \emph{tensor parallelism}~\cite{huang2019gpipe, narayanan2019pipedream, narayanan2021megatron}, divides models into smaller partitions to fit them into per-GPU memory.

\para{Checkpointing and failover.}
Early deep learning training relies solely on data parallelism, and several techniques~\cite{mohan2021checkfreq,nicolae2020deepfreeze,TorchSnapshot} were proposed to speed up checkpointing in this scenario. 
More recent systems like SWIFT~\cite{zhong2023swift}, LightCheck~\cite{chen2023lightcheck}, Gemini~\cite{wang2023gemini}, TRANSOM~\cite{wu2023transom}, Dlrover~\cite{Dlrover}, DataStates-LLM~\cite{maurya2024datastates}, and ByteCheckpoint~\cite{wan2024bytecheckpoint} target LLM training with 3D parallelism. They use asynchronous multi-level checkpointing with various optimizations to saturate disk bandwidth and limit recovery region. 
However, these methods are still far from achieving instant checkpointing because they do not completely eliminate the overhead of checkpointing large and redundant states.
 An alternative, JIT-Checkpointing~\cite{jit-checkpointing}, triggers checkpointing reactively at failure time rather than periodically. While this reduces unnecessary I/O—saving model weights and optimizer states only when a failure is imminent—it can forfeit critical optimizer states because only shards present on healthy ranks can be restored.

\para{Fault tolerance.}
Maintaining training under partial node failures is as important as saving and restoring state. Prior systems pursue elasticity or tolerance in different ways: Varuna leverages spot-instance elasticity~\cite{athlur2022varuna}; Bamboo adds redundancy via duplicate computation~\cite{thorpe2023bamboo}; Oobleck and Recycle use pipeline parallelism with heterogeneous templates to survive failures~\cite{jang2023oobleck, gandhi2024recycle}; the stateless parameter server continues with stale weights/gradients~\cite{cao2024sps}; and nonuniform tensor parallelism resynchronizes parameters across differing TP degrees via gradient resharding~\cite{arfeen2025ntp}. These approaches often incur extra cost (e.g., Bamboo’s duplication, Recycle’s bubble time). \sysname is complementary: even with fault tolerance, newly joined nodes must recover state; accelerating failover shortens idle periods and sustains training efficiency.

\para{State sharding.}
In data parallelism, all devices in the same parallel group hold the same model states, leading to wasted GPU memory. Techniques like ZeRO~\cite{ren2021zero, rajbhandari2020zerodp} and FSDP~\cite{zhao2023fsdp} address this issue by state sharding, where optimizer states, gradients, and even model weights are distributed across devices based on user settings. While state sharding saves memory, it introduces challenges for state recovery, as unique states on each device can be lost if a node fails. \sysname overcomes this challenge and thus supports free state sharding in fast failover.

\para{Communication and computation overlap.}
People have developed many techniques to overlap communication and computation during training. For example, we can execute \emph{activation recomputation}~\cite{jain2020activation, korthikanti2023megatron2} or even forward/backward~\cite{jiang2024megascale} during collective communication.
The key to effective workload overlap is accurately estimating computation time. 
Given the number of parameters on each device ($\varphi$), the computation time of forward and backward is based on the number of floating-point operations on each parameter, batch size $b$, token length $s$, and the GPU FLOPS $C$: 
(1) forward computation time: $\frac{2sb \varphi}{C}$; and
(2) backward computation time: $\frac{4sb \varphi}{C}$.
To overlap communication within one iteration, the total transmission time must not exceed the computation time: $T_c = \frac{6sb \varphi}{C}$.
For example, when saving complete {\ckpt}s to remote storage, current {\ckpt} engines must transfer and persist both weights and optimizer states. The {\ckpt} time overhead is: $T_{ckpt} = \frac{16 \varphi (V+I)}{VI}$ with disk/network bandwidth of $I$ and $V$.
\section{\sysname Design Overview}

\subsection{Why LLM Training Recovers Slowly?}
\label{sec:theory}

We elaborate the failover bottlenecks introduced in Section~\ref{sec:intro}.


\para{High {\ckpt} overhead.}
Despite various optimizations to improve {\ckpt} throughput and meet disk bandwidth constraints, the {\ckpt} frequency remains low (typically every 30 minutes, as reported in MegaScale~\cite{jiang2024megascale} and Meta~\cite{dubey2024llama3.1}). This low frequency is due to several factors:
(1) Large {\ckpt} size: LLM {\ckpt}s are very large, making it challenging to store the entire {\ckpt} in memory or write it to disk within one training iteration. For example, a complete {\ckpt} of LLaMA2-70B exceeds 500 GB.
(2) Data network bottleneck: When saving {\ckpt}s from eight GPUs over a shared storage network to disks, the dedicated storage network (often with smaller bandwidth than the training network) can become a bottleneck.
(3) Frequent GPU memory copies: Full {\ckpt} involves copying extensive data from GPU memory, disrupting ongoing computations. Frequent {\ckpt}, even if asynchronous, significantly lowers training efficiency. In our measurement, running asynchronous {\ckpt} in a background thread increases the iteration length by $7 \times$ due to the GPU-host bandwidth contention.

\begin{table}[tb]
\caption{Possibility of GPU failures on different cluster scales and corresponding MFU loss. $P_x$ denotes the cumulative probability of clusters with $x$ GPUs achieving a given MTBF.}
\label{tab:cdf}
\centering
\footnotesize
\begin{tabular}{|c|c|c|c|}
\hline
\textbf{MTBF(hours)} & \textbf{$P_{16384}$} & \textbf{$P_{65536}$} & \textbf{Relative MFU loss} \\ \hline
3                    & 0.46             & 0.91             & 0.19              \\ \hline
6                    & 0.71             & 0.99             & 0.10              \\ \hline
9                    & 0.84             & 0.99             & 0.06              \\ \hline
12                   & 0.91             & 0.99             & 0.05              \\ \hline
\end{tabular}%
\end{table}

\para{Small MTBF.}
The MTBF for most commodity GPUs is typically 80K $\sim$ 100K hours~\cite{A100}, with some GPU models having even lower MTBFs~\cite{dubey2024llama3.1}. Such MTBF values are adequate for small clusters but insufficient for large LLM training clusters. Table~\ref{tab:cdf} shows failure probabilities at different MTBFs and cluster scales, along with the impact on GPU usage, measured by \emph{model FLOPs utilization} (MFU). Given a {\ckpt} overhead of $T_{ckpt}$ and a {\ckpt} interval of $T_i$, we can compute the relative MFU loss as the sum of losses from three resources: $L_{ckpt}$, $L_{recover}$, and $L_{rollback}$, where $L_{ckpt} = \frac{T_{ckpt}}{T_i+T_{ckpt}}$, $L_{recover}=\frac{MTTR}{MTBF+MTTR}$, and $L_{rollback}=\frac{T_i/2}{MTBF+MTTR}$.

Even considering GPU failures alone, a cluster with tens of thousands of GPUs is likely to experience a breakdown every 3 hours, with a 91\% probability. Moreover, even with frequent {\ckpt} (e.g., every 30 minutes) and no {\ckpt} overhead, a 3-hour breakdown results in a 19\% MFU loss. Network, host, and disk failures only further reduce MTBF.
Therefore, without efficient failover solutions, large-scale LLM training faces substantial MFU loss.

\begin{figure}[tb]
	\centering\includegraphics[width=1.0\columnwidth]{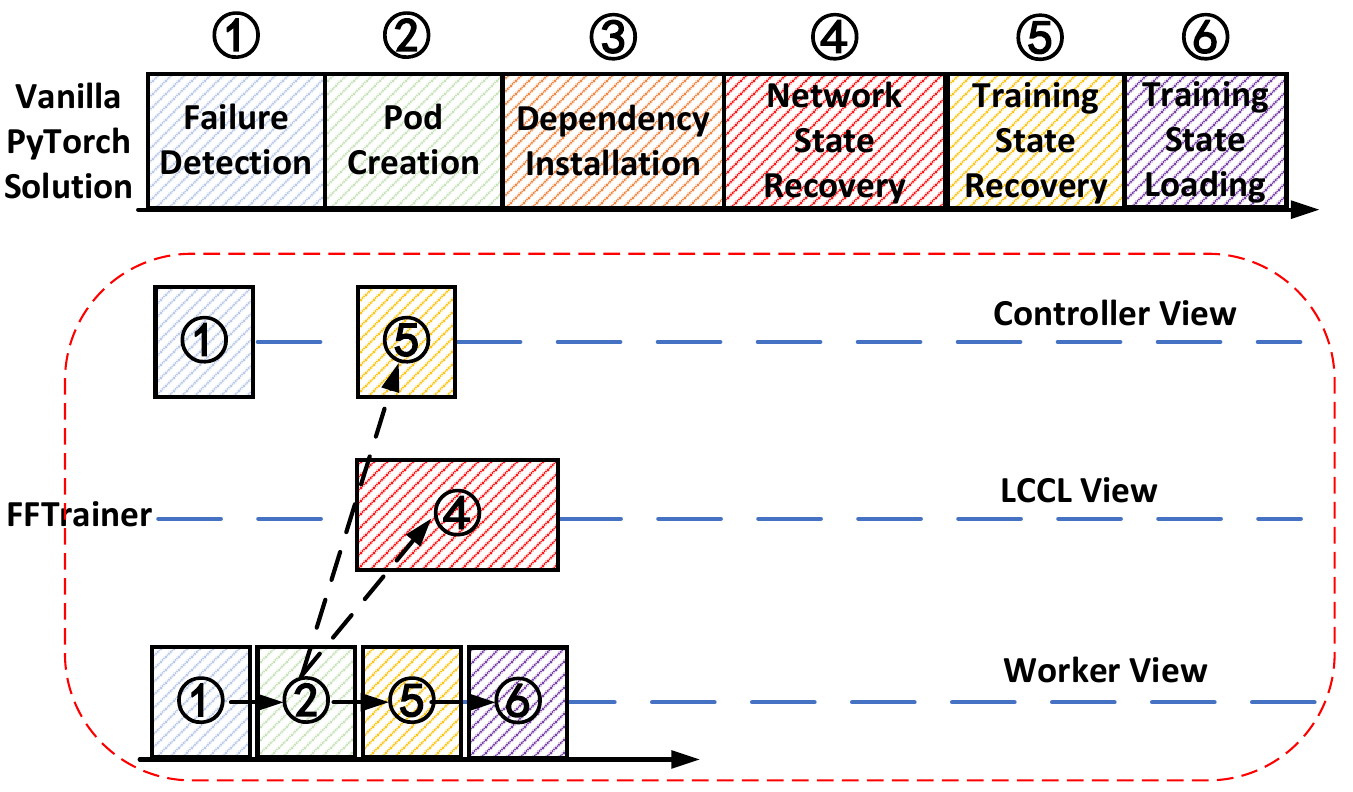}
	\caption{\sysname recovery timeline vs. PyTorch solutions.} 
	\label{fig:motivation}
\end{figure}

\para{High MTTR.}
We measure the overhead of each recovery step in Figure~\ref{fig:motivation} and identify inefficiencies leading to high MTTR:

Step \num{1}: Slow failure detection. When failures occur during blocking collective communication, detecting them involves balancing accuracy with false positives from worker desynchrony. Processes must wait for a communication timeout to exit if peers fail. For instance, NCCL sets default timeout to ten minutes to avoid unnecessary restarts.

Step \num{2} \num{3}: Pod creation and dependency installation. Even with sufficient resources, this step incurs overhead from three tasks: pulling Docker images to the host (e.g., minutes for an NVIDIA PyTorch image); creating and launching the pod (seconds); and installing dependencies (minutes).

Step \num{4}: Network state recovery.  Large jobs must rebuild massive communication groups that take long time to initialize. PyTorch uses a TCP-store to collect socket information and build connections with lock-based concurrency. NCCL bootstraps ranks and initializes groups, which becomes increasingly time-consuming with scale (reaching 1000 seconds with 2048 GPUs~\cite{jiang2024megascale}).

Step \num{5} \num{6}: Recovering and loading training states. New pods stage hundreds of GB per GPU (training data and {\ckpt}s) to local disk/DRAM (minutes).
During recovery, each process then loads its model partition from {\ckpt}s (tens of seconds), but only \emph{after} network state is restored in Step~\num{4}, creating a strict dependency chain.

\subsection{What Can We Do About It}
\label{sec:goals}

Guided by the analysis in \S\ref{sec:theory}, we design \sysname to manage state transfers for faster training recovery while minimizing network and storage costs. It targets two goals: (1) \emph{instant checkpointing}---performing {\ckpt} every iteration with no added overhead---to avoid rollbacks to stale checkpoints, and (2) reduced MTTR for faster recovery. We now introduce LLM-specific optimizations that realize these goals.

\para{[Optimization 1] Reducing {\ckpt} size and transferring them using idle network.  }  
Training progress is lost if the {\ckpt} interval exceeds one iteration. However, previous solutions cannot checkpoint every iteration due to bandwidth/storage limits and large {\ckpt}s.
\sysname enables instant (per-iteration) {\ckpt} via: (1) \emph{checkpoint razor}: eliminates redundant state data, compressing the {\ckpt} size to one-tenth or less; (2) \emph{neighboring redundancy}: backs up {\ckpt}s in their neighbors' memory using idle training network bandwidth from allreduce connections, avoiding data network and disk usage; (3) \emph{lazy backup}: persists and transfers redundant states only during failure recovery.

\para{[Optimization 2] Recovering network states with low latency.} 
PyTorch+NCCL incurs slow communication initialization, often requiring thousands of seconds to derive network states for collectives.
Yet LLM training under ring-style algorithms needs only \emph{two} persistent peers per rank (one for PP, one for DP). \ccl streamlines initialization by discarding heavyweight abstractions and features, establishing minimal point-to-point channels with fast setup for failure recovery.

\para{[Optimization 3] Coordinating multiple types of data transfers within the training network.  } 
\sysname coordinates \traintransfer and \statetransfer traffic, each with distinct optimization goals. \traintransfer includes training output traffic, such as gradients and activations, while \statetransfer involves saving/loading of LLM states like training data and {\ckpt}.
Without careful scheduling, the coexistence of \traintransfer and \statetransfer in the same network can slow down training, as GPU computation depends on inputs from \traintransfer.
To avoid this, \sysname prioritizes \traintransfer to monopolize the network link and complete the transfer quickly, while \statetransfer is allowed only when the network is idle.

\para{[Optimization 4] Speeding up training job restart .  } 
Launching training jobs on a new pod involves several dependent steps: pod creation, dependency installation, network/training state recovery. Existing frameworks execute these steps sequentially, inflating restart latency. \sysname shortens the critical path via:
(1) fully packaged Docker images pre-pulled to hosts, making pods training-ready within seconds;
(2) \ccl-managed inter-node communication with lightweight initialization, reducing connection setup to \emph{tens of seconds} even at scale; and
(3) decoupling training roles from network ranks in \ccl, which lets \sysname overlap model loading with connection establishment.
Figure~\ref{fig:motivation} contrasts \sysname’s overlapped timeline with current serial flows.

\para{Non-goals.}
\sysname is designed to complement existing DT libraries like Megatron and DeepSpeed, remaining transparent to GPU training frameworks and requiring minimal changes to user code. \sysname deliberately does not manage intra-host data paths; it delegates local communication to NCCL and focuses solely on node-level failover.

\subsection{System Architecture}
\label{sec:arch}

\sysname has four modules that jointly support three core functionalities: data management, {\ckpt}, and recovery.

\para{State controller.}
A dedicated process that coordinates data transfer and resource management. At init, it helps \ccl exchange addresses and establish connections. Each iteration, it assigns data indices to data loaders and collects worker heartbeats for liveness. On recovery, it retrieves redundant state from active workers and forwards it to newcomers. Although centralized, it avoids overload by handling only control plane messages.

\para{{\ckpt} engine.} 
\sysname's {\ckpt} engine builds on asynchronous saving to alleviate interruptions during training. It provides interfaces for trainers to save {\ckpt}s for neighboring redundancy or lazy backup.

\para{Data loader.} 
An extension of PyTorch’s loader that preloads training data \emph{over the training network} using indices from the state controller, bypassing local disks and the host data network. It buffers data for future iterations. Each transfer is labeled by {\tid=\texttt{(role, iter)}}, where \texttt{role} is the three-tuple in \S\ref{sec:intro}. The state controller manages all \tid-to-data mappings, and \ccl uses \tid to address states.

\para{\ccl.} 
API-compatible with \texttt{torch.distributed} to minimize code changes. For each collective, \ccl decides whether to use the network: host-agent handled if needed, otherwise offloaded to NCCL. It supports \emph{asynchronous initialization} and \emph{interruptible blocking communication} (\S\ref{sec:ccl}), yielding faster startup and throughput on par with PyTorch.

\para{Agents.}
\sysname adopts a three-layer runtime. Each node runs a pod that (i) launches a \emph{worker agent} to spawn one per-GPU worker (e.g., via \code{torchrun}), monitor health, reap zombies, and restart workers on state-controller signals; and (ii) runs a \emph{\ccl host agent} to handle all traffic in training network via a user-level stack (RDMA-backed).
\section{Resource-Sensitive State Management}
\label{sec:state}

\begin{figure}[tb]
	\centering\includegraphics[width=1.0\columnwidth]{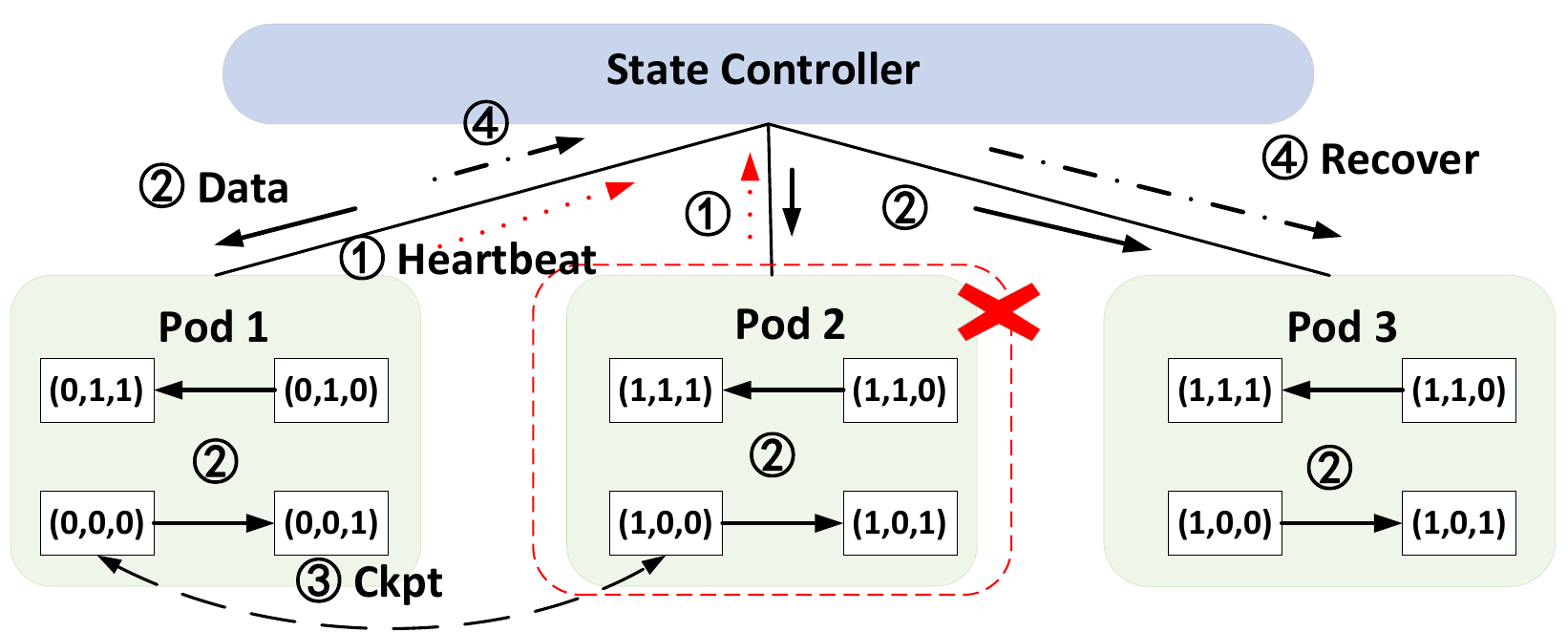}
	\caption{\sysname communication among workers with role $\{r_d,r_p,r_t\}$. Each iteration: workers send heatbeats to the controller (\num{1}) and backup their states on the neighbour in the data parallel group (\num{3}). The controller spreads data indexing to the rank 0 in each tensor parallel group (\num{2}). When pod 2 fails, the controller can detect failure via heartbeats and forward states from pod 1 to the substitute (\num{4}).} 
	\label{fig:flow}
\end{figure}

We require high bandwidth for exchanging gradients and activations, yet links are often underutilized outside bursty collectives. \sysname leverages this unused bandwidth to enable two resource-intensive tasks in distributed training: training data and {\ckpt} management. Figure~\ref{fig:flow} illustrates the communications involved in \statetransfer during normal training and recovery. The key challenge is to support both tasks while maintaining high performance.

\subsection{Training Data Management}
To eliminate data loading stalls, we implement fine-grained data management that delivers training data just-in-time. Key functions include preloading, buffer management, and persistence, ensuring that the training process never waits for data or blocks GPU computation.

Each \sysname data loader maintains a buffer in CPU memory to hold data for the next $k$ iterations (default $k=10$). The buffer size is calculated as the smaller of the $k$ input size and the amount of data that can be transferred in $T_c$: $B=min(4sbk, \frac{6sb\varphi V}{C})$ (about 40 MB for LLaMA-3 70B).

\sysname extends PyTorch's data loader with two key functionalities: (1) preloading and buffering management: an asynchronous thread preloads training data and buffers it in a FIFO manner, evicting used data after each iteration and regulating preloading speed to prevent overflow; and (2) naming resolution: an overridden \code{get\_item} method retrieves data using \tid from the buffer.

In \sysname, workers do not store statically partitioned training data. Instead, the state controller computes and maintains indexing from data items to each \tid, completing this at startup based on the number of GPUs without adding extra overhead. It also tracks how many workers are active and can dynamically adjust batch sizes and indexing.

With the state controller managing data partitions, \sysname allows training jobs to dynamically update sampling algorithms and rearrange data feed order, providing more opportunities to improve training quality.

The indexing information from the state controller decouples the training role from network ranks, allowing \sysname's data loader to preload training data from a \emph{data server} via the training network. The data server, running on globally shared storage or local disks, preloads data for faster response to preloading requests.

\subsection{Checkpointing}
\label{sec:checkpointing}
\begin{figure}[tb]
	\centering\includegraphics[width=1.0\columnwidth]{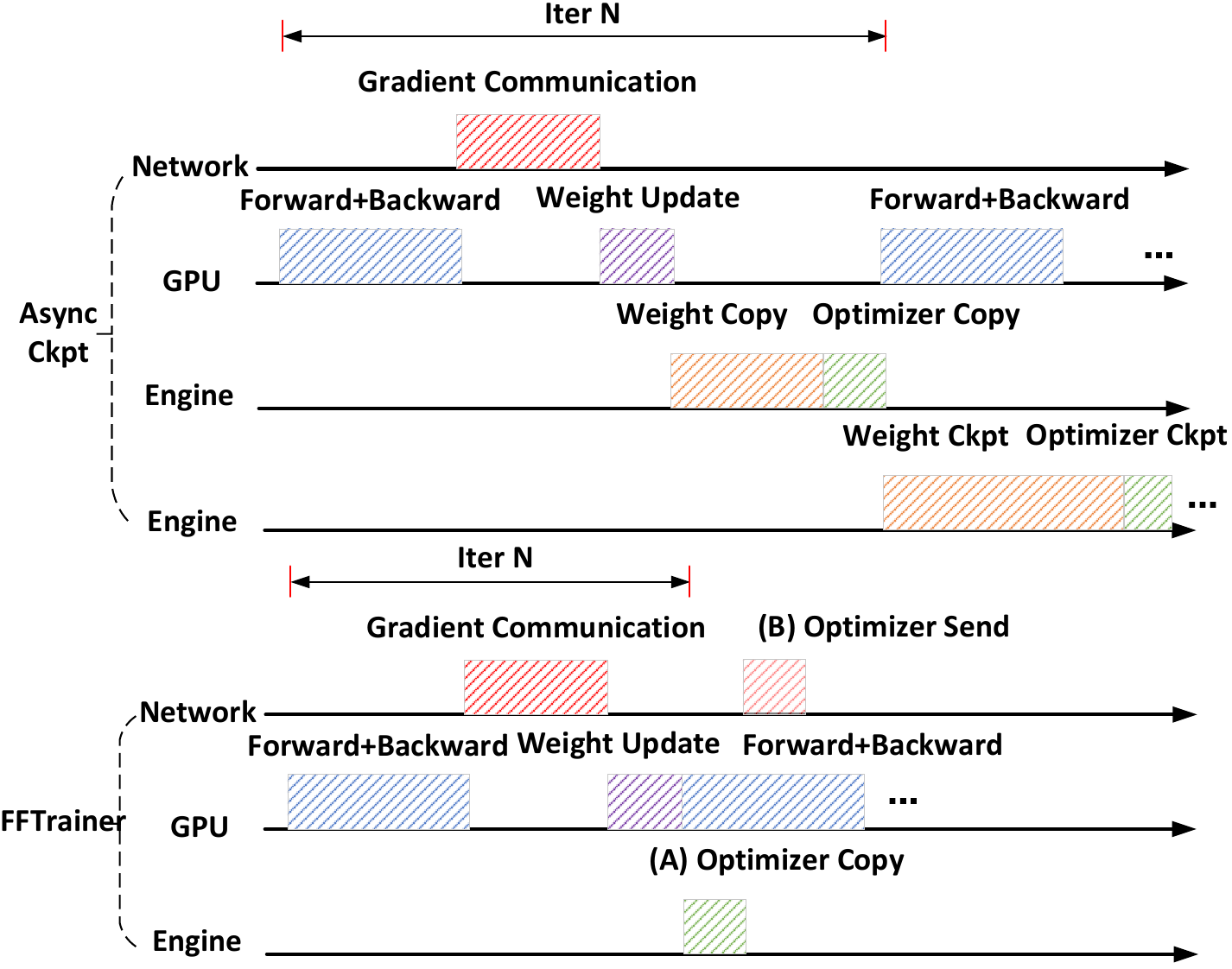}
	\caption{\sysname checkpointing approach vs. current solutions. \sysname introduces no blocking time between iterations but overlaps all checkpointing steps with computation.} 
	\label{fig:ckpt}
\end{figure}

\sysname tailors its {\ckpt} engine to two objectives: (1) minimize interference with training compute, and (2) ensuring that all {\ckpt} steps (both synchronous and asynchronous) are completed within a single iteration. We first introduce the checkpointing techniques and then explain how these techniques enable instant checkpointing. Figure~\ref{fig:ckpt} compares the \sysname's {\ckpt} process with existing frameworks.

\para{Checkpoint razor and fast snapshot.} 
When \sysname's {\ckpt} engine initializes, it parses the training configuration to identify unique states using a tool called the \emph{checkpoint razor}. For LLM training, the checkpoint razor follows two rules: (1) if the data parallel degree is greater than one, model weights are redundant; (2) if the data parallel degree is greater than one and the distributed optimizer is not enabled (as in Zero-2~\cite{ren2021zero} and Megatron~\cite{narayanan2021megatron}), optimizer states are redundant. The checkpoint razor eliminates these redundant states, leaving only unique parts for backup. We save redundant states in healthy workers in {\ckpt} only during job recovery.

The {\ckpt} engine calculates the CPU memory buffer size needed to store unique states based on the optimizer type (e.g., $12\varphi/d$ for the Adam optimizer, where $\varphi$ is the number of parameters per device and $d$ is the data parallel degree).

In each iteration, \sysname snapshots these unique states by device-to-host copies, stores them in a pre-allocated RDMA memory buffer ((A) in Figure~\ref{fig:ckpt}), and streams them to the next worker's buffer in the data parallel ring (i.e., neighboring redundancy, (B) in Figure~\ref{fig:ckpt}). 
By avoiding disk writes and focusing only on unique states (typically small due to data parallelism), \sysname accelerates the snapshot process. It also avoids serializing states with \code{Pickle}~\cite{pickle}, further reducing overhead. These optimizations allow \sysname to achieve instant checkpointing.

We model the relationship between a model's computation overhead and the transfer overhead of its optimizer states,  exploring the conditions required to overlap transfer with computation. The following inequality captures the condition:
\begin{equation}
	\label{eq:ratio1}
	T_c = \frac{6sb\varphi}{C} \geq T'_{ckpt} = \frac{12\varphi}{V}.
\end{equation}
Using the checkpoint razor, we reduce the {\ckpt} time from $T_{ckpt}=\frac{16 \varphi (V+I)}{VI}$ to $T'_{ckpt} = \frac{12\varphi}{V}$, achieving over a 90\% reduction.
We also find that as long as the ratio of the model size and NIC bandwidth to GPU FLOPS is below a threshold (i.e., $\frac{sbV}{2C} \geq 1$), \sysname can fully hide the neighboring redundancy process within the computation. 
We call $\frac{sbV}{2C}$ the \emph{free checkpointing ratio (FCR)}, a metric indicating whether {\ckpt} overhead can be hidden within the computation, and we have the FCR condition:
\begin{equation}
	\label{eq:ratio2}
    T_c \geq T'_{ckpt} \quad \text{iff} \quad  \text{FCR}=\frac{sbV}{2C} \geq 1.
\end{equation}
In Section~\ref{sec:evaluation}, we will show that this condition is satisfied in most real cases. 

\para{Lazy backup and checkpointing consistency. }
Recovery requires both unique and redundant state. \sysname adopts \emph{lazy backup}: upon restart, healthy workers persist only the redundant state \emph{before} exiting; to minimize duplication and ensure consistency, only \code{rank 0} in each DP group performs this backup. The backup runs in parallel with new-pod creation, adding negligible restart overhead and remaining compatible with future deployments.

Because \sysname does not enforce a fixed {\ckpt} iteration, failures that stall collectives can leave groups at different versions (e.g., some checkpoints update in iteration $n$, others in $n+1$). The state controller resolves this by choosing the earliest available iteration as the global {\ckpt} and instructing survivors to roll back to that version for lazy backup. Model weights can be reconciled using the latest gradients, but optimizer state is harder; therefore, \sysname keeps two recent snapshots of optimizer state for version coordination—typically a few GB that fits in CPU memory.

\para{Corner case handling for correctness. }
Two corner cases can invalidate {\ckpt}s in \sysname: (1) all workers in a data parallel group fail, and (2) a worker fails along with its next rank. While increasing redundancy by saving states across more workers could address these issues, it is not worth the overhead to handle these rare cases (approximately 1.7\% during recovery, as Table~\ref{tab:cdf} shows). To avoid adding system complexity and overhead, \sysname employs multi-level insurance for these rare events: after a long training period (e.g., 500 iterations), \sysname triggers traditional asynchronous {\ckpt} to save complete states, ensuring a fallback mechanism for basic recovery even if instant checkpointing fails.

\subsection{Load Planning}
\label{sec:load}
\para{Reducing workload of state controllers.}
\sysname employs a single controller per job—even at scale—and confines it to lightweight control-plane duties. To ensure scalability, it limits data I/O and message volume as follows.
First, only the local \code{rank 0} on each pod reports per second, capping controller connections at $1/8$ of total GPUs. States are updated in a lock-free array, enabling failure detection within one second even with tens of thousands of GPUs.
Second, because workers in a model-parallel group share the same data, the controller sends indices only to that group's \code{rank 0}; intra-node tensor-parallel links fan out the indices to peers. Even at large scales, the controller transmits less than one GB per iteration, taking under a second using the training network.

\para{Fast pod creation.}
\sysname employs additional operating techniques to reduce restart overhead. When the job first starts, \sysname pulls the required Docker image to all available hosts and maintains this list throughout the job's runtime. This optimization ensures new pods created for recovery do not waste time pulling images. Additionally, \sysname pre-installs all training dependencies in a pod and commits it as a new Docker image before launching the job, avoiding repeated downloads and installations on each pod.
\section{Latency-Sensitive State Management}
\label{sec:ccl}


\sysname uses \ccl to overcome NCCL's high latency in network state recovery by several techniques. 
We describe these techniques in this section.

\subsection{Fast Network State Recovery}
The default DT setup incurs significant overhead from \emph{connection building} and \emph{communication group initialization}. In PyTorch, TCP-Store~\cite{tcpstore} exchanges network addresses and synchronizes communication group initialization on a single thread, causing the read-write overhead to grow linearly with the number of GPUs. Moreover, 3D parallelism requires multiple groups per worker, resulting in $O(N^2)$ initialization complexity due to global barriers. \ccl addresses these inefficiencies with two techniques:

\para{Group-free collective communication.}
\sysname customizes collective communication for ring-based 3D parallelism. In distributed training, each worker needs at most four inter-node connections—two for DP and two for PP. With static participation, MPI-style membership management is unnecessary. Thus, \sysname eliminates cross-node communication groups in \ccl, bypassing costly group initialization. Workers simply send data through fixed connections.

\para{Lock-free connection building.}
The state controller uses a lock-free array for asynchronous exchange of node addresses. Each node writes its rank and address and retrieves two target addresses for its receivers. By flagging each address upon completion, threads can verify readiness, allowing workers to build connections without barrier synchronization. This process is repeated during failover in \sysname, ensuring that a node with an LCCL agent can always track the mapping between its global role and rank, even after a failure occurs.

\subsection{Overlapping Initialization with Model Loading}
\emph{Model loading} from disk to GPU is one of the most time-consuming steps in training initialization. However, existing frameworks typically execute the step sequentially after communication initialization because each device must identify its ranks to determine which model partition to load (Figure~\ref{fig:motivation} in Section~\ref{sec:theory}). Both steps are slow in LLM training, further delaying restarts.
\ccl introduces an asynchronous two-stage initialization to overlap these steps. First, it uses PyTorch interfaces to establish GPU connections within a single host via NCCL, which completes quickly (within hundreds of milliseconds) as it only involves 8 GPUs within a node. Meanwhile, the host agent can derive a set of roles from the state controller. Roles are then assigned to each worker according to their local ranks, allowing model loading to begin while \ccl builds cross-node connections simultaneously.

Thanks to the initialization speedups above, connection building usually completes faster than model loading. \ccl ensures that collective communication is delayed until all initialization is finished, guaranteeing correctness in rare cases.

\subsection{Guaranteeing Communication Performance} 
\ccl offloads intra-node communication to NCCL, utilizing its high-performance implementation on PCIe/NVlink~\cite{NVLink}. For multi-node allreduce operations, \ccl performs an intra-node reduce followed by a ring allreduce among host agents to compute the final results. After cross-node communication, the results are broadcast within the host via NCCL. This approach achieves performance on par with PyTorch using NCCL, benefiting from high intra-node bandwidth.

To ensure consistent user interfaces, \ccl uses \emph{faked groups}. For instance, when users create a faked group $[0,4,8,12]$ across two hosts, \ccl internally builds two local process groups, $[0,4]$ and $[8,12]$, in NCCL and only manages the data exchange across groups.

To optimize network usage, \ccl allows \traintransfer to monopolize the network, with \statetransfer occurring only when no gradient or activation transfers are in progress. This is achieved by maintaining separate queues for \traintransfer and \statetransfer. 
Specifically, data loaders preload training data only when buffer space is available and the downlink is idle, ensuring data preloading does not block training. The checkpoint engine follows a similar approach: checkpoint threads back up in-memory states to neighboring workers only when there are no ongoing activation or gradient transfers.

As noted in Table~\ref{tab:data} in Section~\ref{sec:intro}, training data bandwidth requirements typically remain below one megabyte per iteration for most models, leaving sufficient spare bandwidth for all workers, even on hosts with eight GPUs and a single NIC.

\section{Fast Failure Detection and Failover}
\sysname assumes the fail-stop model commonly adopted in large-scale training. In this section, we discuss how \sysname detect and recover from these failures.

\begin{table}[t]
\caption{\sysname interacts with external modules in different cases during failover.}
\label{tab:recovery}
\centering
\resizebox{\columnwidth}{!}{%
\begin{tabular}{|c|ccc|c|}
\hline
\multirow{2}{*}{}                                                        & \multicolumn{3}{c|}{\textbf{External}}                                                                                                                                                                                                                      & \textbf{Internal}                                                            \\ \cline{2-5} 
                                                                         & \multicolumn{1}{c|}{\textbf{Operator}}                                                  & \multicolumn{1}{c|}{\textbf{Pod}}                                              & \textbf{Agent}                                                                   & \textbf{FFTrainer}                                                           \\ \hline
\textbf{\begin{tabular}[c]{@{}c@{}}Normal \\ launch\end{tabular}}        & \multicolumn{1}{c|}{\begin{tabular}[c]{@{}c@{}}create \\ pods\end{tabular}}             & \multicolumn{1}{c|}{\begin{tabular}[c]{@{}c@{}}create an \\ agent\end{tabular}} & \begin{tabular}[c]{@{}c@{}}create \\ workers\end{tabular}                        & N/A                                                                          \\ \hline
\textbf{\begin{tabular}[c]{@{}c@{}}Healthy node \\ restart\end{tabular}} & \multicolumn{1}{c|}{N/A}                                                                & \multicolumn{1}{c|}{N/A}                                                       & \begin{tabular}[c]{@{}c@{}}restart \\ workers\end{tabular}                       & \begin{tabular}[c]{@{}c@{}}failure \\ detection; \\ lazy backup\end{tabular} \\ \hline
\textbf{\begin{tabular}[c]{@{}c@{}}Software \\ failure\end{tabular}}     & \multicolumn{1}{c|}{N/A}                                                                & \multicolumn{1}{c|}{N/A}                                                       & \begin{tabular}[c]{@{}c@{}}failure \\ detection; \\ restart workers\end{tabular} & \begin{tabular}[c]{@{}c@{}}recover \\ {\ckpt}\end{tabular}                      \\ \hline
\textbf{\begin{tabular}[c]{@{}c@{}}Hardware \\ failure\end{tabular}}     & \multicolumn{1}{c|}{\begin{tabular}[c]{@{}c@{}}reschedule \\ failure pods\end{tabular}} & \multicolumn{1}{c|}{\begin{tabular}[c]{@{}c@{}}create an \\ agent\end{tabular}} & \begin{tabular}[c]{@{}c@{}}create \\ workers\end{tabular}                        & \begin{tabular}[c]{@{}c@{}}recover \\ {\ckpt}\end{tabular}                      \\ \hline
\end{tabular}%
}
\end{table}

\subsection{Cross-layer Worker Failure Detection}
Blocking collective communication is essential for training, as workers rely on it to synchronize their steps, adding \emph{barrier} semantics beyond simple data exchange. 
As we discuss in Section~\ref{sec:theory}, tolerating the collective communication timeout to exit normally is time-consuming.
Prior solutions use a worker agent to detect failures and forcefully terminate workers. However, this approach results in the loss of all active states in healthy pods, violating the lazy checkpointing policy and leading to a longer detection period.

To address this issue, \sysname employs a cross-layer signal to implement blocking communication in a non-blocking manner. When a collective communication operation is invoked, \sysname first submits the request to the host agent and then waits for signals from either communication completion or interruption. Upon completion, the host agent receives the data and notifies all waiting workers to proceed. 

If a failure occurs during the communication, the state controller detects missing application-level heartbeats and triggers a breakdown notification to host agents, prompting a job recovery. Host agents wake up waiting workers and allow worker agents to restart unresponsive processes.
This method is much faster than waiting for communication timeouts, as heartbeats are sent frequently (by default, every second). It is also more accurate than external heartbeats because the training process itself reports real-time aliveness via \ccl.

The hybrid signal design for communication interruption has two key benefits in implementation: (1) waiting threads can release the \emph{global interpreter lock (GIL)} of Python, allowing other threads to receive messages from the state controller, and (2) the main training thread can exit normally, enabling lazy backup before the restart.

\subsection{Failover}
\label{sec:failover}

\sysname uses the above mechanisms to handle different failure cases. 
Table~\ref{tab:recovery} enumerates all roles and their actions on different training recovery cases:
(a) Normal launch: the operator creates a pod on each node; the launch command of each pod creates a worker agent; the agent then creates a worker for each GPU; (b) Healthy node restart: workers receive notifications from the state controller for lazy backup and then exit normally; the agent creates new workers; (c) Software failure: the worker agent receives notifications from the state controller to restart some workers and then workers recover from  {\ckpt}s; (d) Hardware failure: the operator kills the original pod and repeats (a) on a new node, and workers additionally recover states from other nodes.
Different cases share the same steps in Section~\ref{sec:theory} after workers restart.

We analyze the recovery probability of \sysname under a common failure assumption where, in a cluster of $N$ machines, exactly $k$ machines fail. The recovery probability from the {\ckpt} in main memory is given by
\begin{equation}
	\label{eq:ratio3}
\left\{
\begin{aligned}
&P_r(N,k)=1, &\quad \text{if } k = 1, \\
&P_r(N,k)=\frac{\binom{N-k}{k}+\binom{N-k-1}{k-1}}{\binom{N}{k}},  &\quad \text{if } k > 1.
\end{aligned}
\right.
\end{equation}
For $k>1$, the probability is the fraction of ways to choose $k$ failed servers that include at least one pair of adjacent servers (which causes backup loss) out of all possible ways to pick $k$ failed servers.  
We also consider the probability of such $k$-node failures. Given the MTBF of a single GPU $T_b$, the probability that a host with eight GPUs fails within $H$ hours is $1-e^{\frac{-8H}{T_b}}$. Thus, the probability of $k$-node failures in a cluster with $N$ machines is
\begin{equation}
	\label{eq:ratio4}
P_f(N,k,H) = \binom{N}{k}(1-e^{-\mu H})^k(e^{-\mu H})^{N-K}, 
\end{equation}
where $\mu = \frac{8}{T_b}$.
The overall probability of successful recovery is
\begin{equation}
	\label{eq:ratio5}
P(N,H) = \sum_{k=0}^N P_r(N,k) \times P_f(N,k,H).
\end{equation}
$P(N,H)$ is large enough for most training scenarios (higher than 99\% within 12 hours even when $N \geq 10,000$).

As noted in Section~\ref{sec:checkpointing}, all workers must restart from the same iteration; if some are ahead, we roll them back to the most recent globally consistent iteration. This rollback does not reduce training progress because we checkpoint the optimizer state immediately after each update (i.e., before the start of the next iteration). Consequently, once all workers have completed the weight update and entered the next iteration, a complete optimizer-state \ckpt exists, and training can safely resume from that iteration.
\section{Evaluation}
\label{sec:evaluation}

We evaluate \sysname to answer three questions: (1) its overhead and MTTR improvements on real clusters; (2) the source of these improvements; and (3) its scalability to larger scale and future faster clusters.  We answer the first two using experiments on a real cluster and the third with analysis. 

\subsection{Experiment Setup}
\begin{table}[t]
\caption{Default training configurations and models used in experiments.}
\label{tab:param}
\centering
\resizebox{\columnwidth}{!}{%
\begin{tabular}{|c|c|c|c|c|}
\hline
           & \textbf{GPT-2 2.7B} & \textbf{LLaMA3-8B} & \textbf{LLaMA2-13B} & \textbf{LLaMA3-70B} \\ \hline
\textbf{$d$}          & 16     & 4         & 4          & 2          \\ \hline
\textbf{$p$}         & 2     & 8         & 8          & 8          \\ \hline
\textbf{$t$}          & 4     & 4         & 4          & 8          \\ \hline
\textbf{Batch size} & 512    & 256        & 256         & 128        \\ \hline
\end{tabular}%
}
\end{table}

\para{Testbed.} We run \sysname on a testbed of 128 GPUs on 16 servers connected via 200 Gbps Infiniband. 
Each worker node has a single 200 Gbps Mellanox ConnectX-6 NIC, $2\times$ 56-core 2.00GHz CPUs, 192 GB RAM, and $8\times$ NVIDIA GeForce RTX 4090 GPUs. All nodes use NVIDIA driver 550.76 with CUDA 12.4, Mellanox OFED 5.8-3.0.7.0, and Ubuntu 20.04. 

\para{Prototype implementation.  }  We integrate \sysname with PyTorch, DeepSpeed, and Megatron as four modules: two plugins that implement interfaces to \ccl and the {\ckpt} engines on PyTorch, each with 1K lines of Python; \ccl with 4K lines of C++; and a state controller with 1K lines of Python and 1K lines of C++.

\para{Workloads.} We use four LLMs with various sizes: GPT-2 2.7B, LLaMA3-8B, LLaMA2-13B, and LLaMA3-70B.  We perform pre-training using the Common Crawl dataset~\cite{CommonCrawl}.  We train the models without quantization using fp16.  We choose the tensor/pipeline parallel parameter $t$ and $p$ so the training fits the GPU memory, then we fully utilize all 128 GPUs by setting the correct data parallel degree $d$. Table~\ref{tab:param} summarizes the configurations.


\para{Baselines.}
We compare \sysname with the original checkpoint engines from Megatron~\cite{narayanan2021megatron}, DeepSpeed~\cite{DeepSpeed}, and PyTorch~\cite{TorchSnapshot}. In addition, we evaluate state-of-the-art failover solutions such as MegaScale~\cite{jiang2024megascale}, which focuses on fast recovery, and Gemini~\cite{wang2023gemini}, which emphasizes rapid {\ckpt}.



\subsection{Key performance results}
\begin{figure}[t]
	\centering\includegraphics[width=1.0\columnwidth]{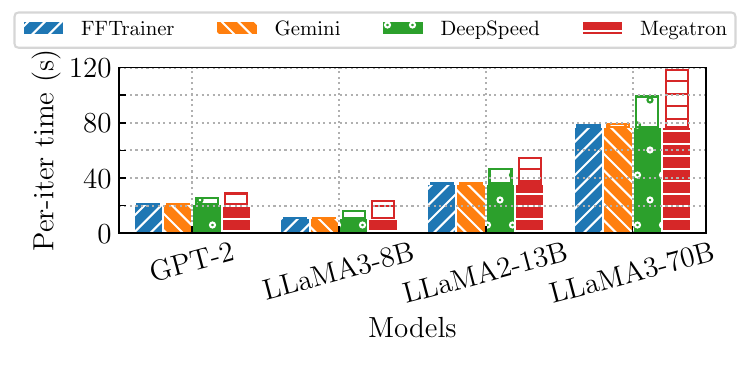}
	\caption{Checkpointing overhead for each system.  The hollow part represents the difference of training time with/without {\ckpt}, i.e., overhead of asynchronous checkpointing.} 
	\label{fig:iter-time}
\end{figure}

\para{\sysname achieves instant checkpointing with minimal impact on training performance. }
We measure the per-iteration time of different models using various {\ckpt} engines, and show the performance comparison in Figure~\ref{fig:iter-time}.  The {\ckpt} overhead is calculated by running the training with/without {\ckpt}, and taking the difference (shown as the hollow part of the bars). 
While \sysname does {\ckpt} every iteration, other engines fail to support this high frequency, and thus we only do {\ckpt} every five iterations on vanilla DeepSpeed and Megatron. Gemini does {\ckpt} every minute as in its evaluation.   

We observe that:
(1) Despite lower {\ckpt} frequency, DeepSpeed and Megatron increase training time by 23\% to 110\% due to {\ckpt} traffic, which can last several iterations (Section~\ref{sec:theory}).
(2) As model size grows, the {\ckpt} overhead increases for baselines due to the larger {\ckpt} size.
(3) In contrast, \sysname maintains an overhead of less than 3\% in all cases, demonstrating the benefits of the checkpoint razor and neighboring redundancy (Section~\ref{sec:state}).
(4) Although both \sysname and Gemini incur minimal {\ckpt} overhead during training, \sysname checkpoints every iteration while Gemini checkpoints every minute (about 1 to 6 iterations), allowing \sysname to achieve a higher {\ckpt} frequency.

\begin{table}[t]
	\caption{Breakdown of failover overhead (seconds) when training LLaMA2-13B.}
	\label{tab:failover}
	\centering
 \resizebox{\columnwidth}{!}{%
	\begin{tabular}{|c|cc|cc|c|}
\hline
                                     & \multicolumn{2}{c|}{\textbf{Gemini}}             & \multicolumn{2}{c|}{\textbf{\sysname}} & \multirow{2}{*}{\textbf{\begin{tabular}[c]{@{}c@{}}Time \\ Reduction\end{tabular}}} \\ \cline{1-5}
\textbf{GPU num}                     & \multicolumn{1}{c|}{16}           & 128           & \multicolumn{1}{c|}{16}    & 128    &                                                                                    \\ \hline
\textbf{Failure detection}           & \multicolumn{1}{c|}{15}  & 15   & \multicolumn{1}{c|}{6}     & 6      & 60\%                                                                               \\ \hline
\textbf{Pod creation}                & \multicolumn{1}{c|}{392}   & 392    & \multicolumn{1}{c|}{7}     & 7      & 98\%                                                                               \\ \hline
\textbf{Dependency install}          & \multicolumn{1}{c|}{421}   & 421    & \multicolumn{1}{c|}{0}     & 0      & 100\%                                                                              \\ \hline
\textbf{\begin{tabular}[c]{@{}c@{}}State recovery \\ \&\& loading\end{tabular} } & \multicolumn{1}{c|}{71}           & 166           & \multicolumn{1}{c|}{13}    & 16     & 82-91\%                                                                               \\ \hline
\textbf{Total}                       & \multicolumn{1}{c|}{899} & 994 & \multicolumn{1}{c|}{26}    & 29     & 97\%                                                                               \\ \hline
\end{tabular}%
	}
\end{table}

\para{\sysname reduces MTTR at every step of the recovery.  }
Referring to the six steps illustrated in Figure~\ref{fig:motivation} (Section~\ref{sec:goals}), we measure the time spent on each step, using LLaMA2-13B as an example.  Note that these numbers only depend on per-worker state size but not the total model size, and thus all models provide similar results.  
Table~\ref{tab:failover} summarizes the results, and our observations are as follows: 
(1) By avoiding collective communication timeout, \sysname reduces detection time to 6 seconds.
(2) Pre-installing and pre-loading all pod images allows \sysname to start pods within 7 seconds.
(3) Decoupling network and training initialization enables both steps, including network/training state recovery and training state loading, to be completed within 16 seconds.
(4) When increasing from 16 to 128 GPUs, connection building time increases by only $27\%$ with \sysname, compared to an increase of $2.3\times$ with the off-the-shelf PyTorch implementation. 
In summary, \sysname can reduce the MTTR  by $97\%$.  

\begin{figure}[t]
	\centering\includegraphics[width=1.0\columnwidth]{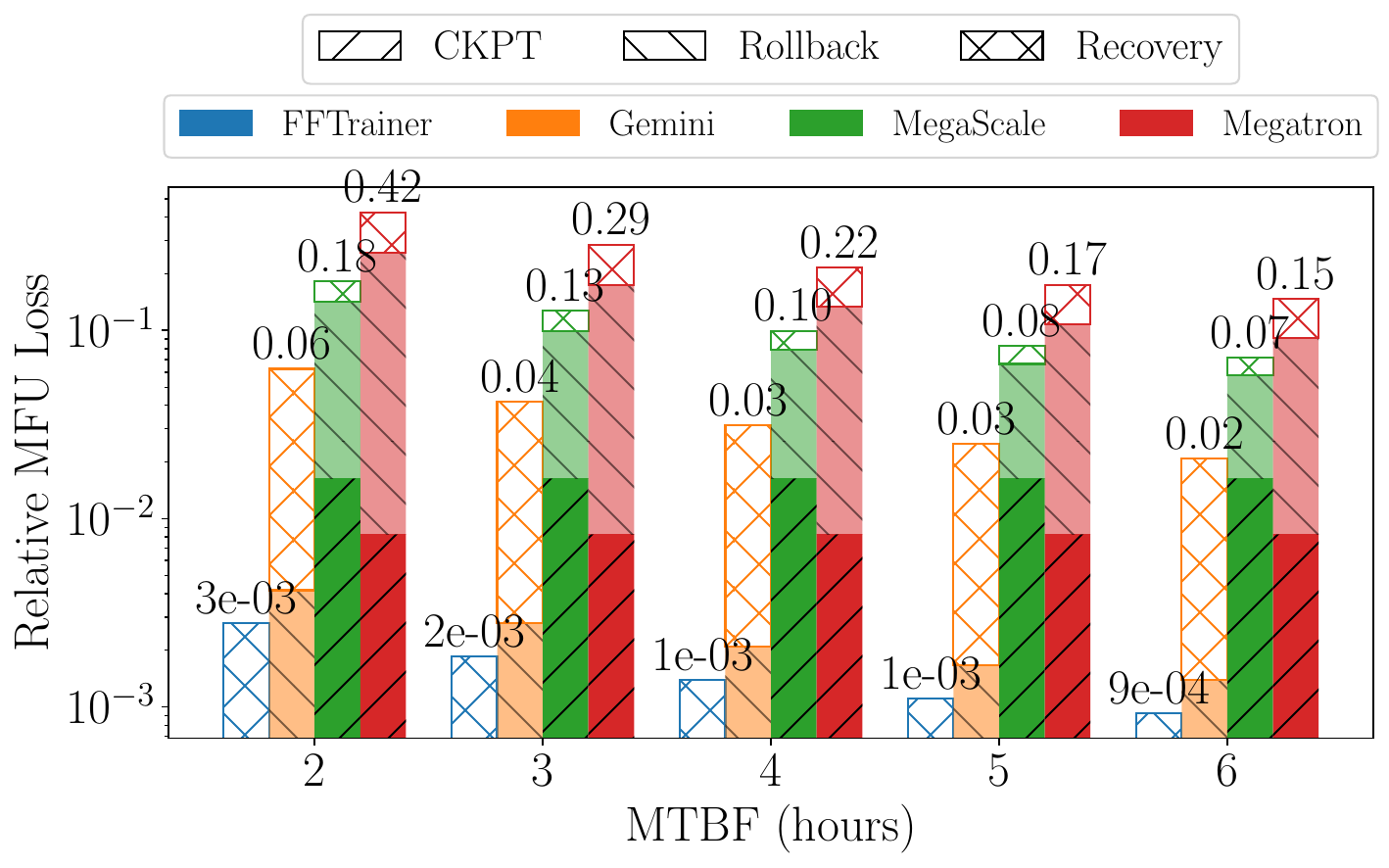}
	\caption{Relative MFU loss for failures at different MTBF.} 
	\label{fig:mfu}
\end{figure}

\para{Combining frequent {\ckpt} with reduced MTTR significantly improves GPU utilization.  } 
We use MFU as the effectiveness metric for failover mechanisms in log scale. Using the measurement results from \sysname, Megatron, Gemini, and MegaScale, Figure~\ref{fig:mfu} plots the MFU loss with the cluster MTBF ranging from 2 to 6 hours. The {\ckpt} intervals for the four systems are per-iteration, per-minute, per-half-hour, and per-hour, as suggested in their respective papers.

We observe the following:
(1) For Megatron, too frequent {\ckpt} leads to significant checkpoint loss (solid bars), while infrequent checkpoints cause substantial progress loss due to rollback to previous {\ckpt}  (semitransparent bars).
(2) Although Gemini reduces {\ckpt} overhead, it still suffers from rollback (about 1\%) and recovery (about 5\%) overhead in short-MTBF cases without fast failover optimizations.
(3) MegaScale speeds up recovery but is limited by its {\ckpt} frequencies and lacks checkpoint optimizations, resulting in 12\% rollback loss (semitransparent bars).
(4) \sysname reduces MFU loss to nearly zero (no more than 0.27\%) by combining frequent, efficient {\ckpt} with reduced MTTR.
(5) Since MTTR is constant, when MTBF is small, the MFU loss due to recovery becomes more significant (hollow bars).

\subsection{Deeper look into each component}



\begin{table}[t]
\centering
\caption{The probability that \sysname can recover failures from {\ckpt}s stored in main memory ($P(N,H)$).}
\label{tab:prob}
\resizebox{\columnwidth}{!}{%
\begin{tabular}{|c|c|c|c|c|}
\hline
\begin{tabular}[c]{@{}c@{}}\textbf{number of hosts} \\ \textbf{(8 GPUs each)}\end{tabular} & \begin{tabular}[c]{@{}c@{}}\textbf{FFTrainer}\\ \textbf{H=3}\end{tabular} & \begin{tabular}[c]{@{}c@{}}\textbf{FFTrainer}\\ \textbf{H=12}\end{tabular} & \begin{tabular}[c]{@{}c@{}}\textbf{Gemini}\\ \textbf{m=2,H=3},\end{tabular} & \begin{tabular}[c]{@{}c@{}}\textbf{Gemini}\\ \textbf{m=2,H=12}\end{tabular} \\ \hline
800                                                                         & 99.98\%                                                 & 99.74\%                                                  & 100.00\%                                                  & 99.94\%                                                   \\ \hline
1200                                                                        & 99.97\%                                                 & 99.65\%                                                  & 99.99\%                                                   & 99.91\%                                                   \\ \hline
1600                                                                        & 99.96\%                                                 & 99.57\%                                                  & 99.99\%                                                   & 99.88\%                                                   \\ \hline
2000                                                                        & 99.96\%                                                 & 99.50\%                                                  & 99.99\%                                                   & 99.86\%                                                   \\ \hline
\end{tabular}%
}
\end{table}

\para{Overall recovery probability.}
To validate the effectiveness of \sysname’s {\ckpt} strategy, we compare its CPU-memory recovery probability with Gemini (Section~\ref{sec:failover}). Table~\ref{tab:prob} plots the recovery probability for both systems as a function of the number of machines. Even in a cluster with 2,000 servers (16,000 GPUs), \sysname’s lightweight {\ckpt} strategy achieves a success probability nearly identical to Gemini, with less than a 0.5\% difference over 12 hours. In recovery failure cases, both systems load {\ckpt}s from remote storage, incurring higher latency. These results validate that a single {\ckpt} suffices to guarantee high recovery success while reducing bandwidth and main memory overhead.

\begin{table}[t]
\caption{Impact of \sysname on training efficiency with different parallel degrees when training GPT-2 2.7B.}
\label{tab:parallel}
\centering
\footnotesize
\begin{tabular}{|c|c|c|c|c|}
\hline
                           & \textbf{(2,4,2)} & \textbf{(4,4,2)} & \textbf{(8,4,2)} & \textbf{(16,4,2)} \\ \hline
\textbf{Per-iter time (s)}     & 7.7              & 11               & 12             & 21                \\ \hline
\textbf{\sysname slowdown}    & 1.3\%            & 0.9\%            & 0.8\%            & 1.0\%             \\ \hline
\textbf{Megatron slowdown} & 18\%             & 21\%             & 31\%             & 44\%              \\ \hline
\end{tabular}%
\end{table}

\para{\sysname works well in different parallel configurations.}
We measure the extra cost caused by different {\ckpt} engines at different parallel configurations in Table~\ref{tab:parallel}. In this experiment, we fix the model partition size and batch size on each GPU to keep computation and copy overhead unchanged.
We can see that (1) a larger data parallel degree can further increase {\ckpt} overhead: Megatron {\ckpt} engine spends up to $6.6\times$ more time gathering distributed optimizer states (if enabled) when $d=16$; (2) \sysname introduces little overhead in all cases even with a large data parallel degree.

\begin{figure}[t]
	\centering\includegraphics[width=1.0\columnwidth]{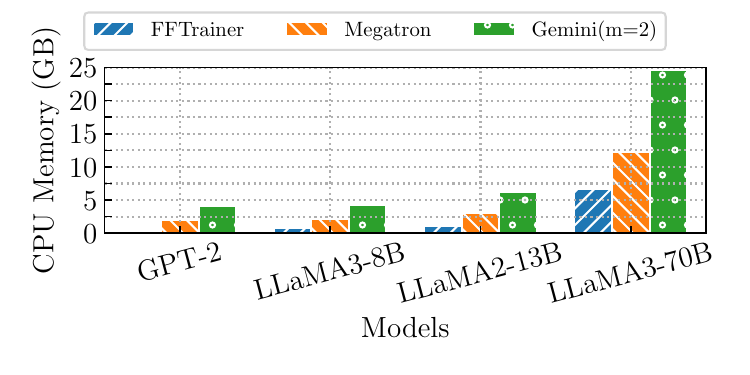}
	\caption{Main memory overhead for checkpointing when training different models.} 
	\label{fig:memory}
\end{figure}

\para{\sysname consumes acceptable main memory.}
A key trade-off in \sysname's design is the use of extra main memory to store {\ckpt}s for subsequent recovery. We measure the main memory overhead of different systems during training, as shown in Figure~\ref{fig:memory}. For Gemini, we set $m=2$ as in its evaluation. \sysname consumes the least main memory compared to all baselines, using 38\% of Megatron and 19\% of Gemini when training LlaMA3-8B. Thanks to the state sharding techniques discussed in Section~\ref{sec:background}, the optimizer state size decreases as the data parallel degree grows, giving \sysname a greater advantage in large-scale training. This benefit is primarily due to \sysname's checkpoint razor optimization, which eliminates redundant training states. Note that the bandwidth consumption on both the network and disk is also equivalent to this overhead, as systems must eventually store {\ckpt}s on remote storage.

\begin{figure}[t]
	\centering\includegraphics[width=1.0\columnwidth]{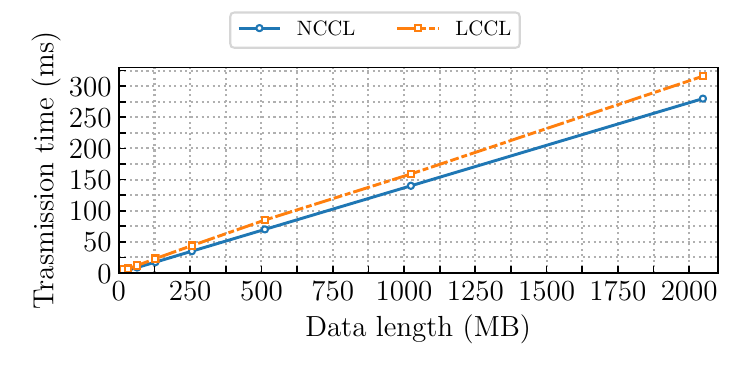}
	\caption{Allreduce time overhead: \ccl vs. NCCL for different data lengths.} 
	\label{fig:ccl}
\end{figure}

\para{\ccl can complete cross-node collective communication efficiently.}
We measure the time overhead of allreduce operations between two physical nodes with increasing data length when using different backends in Figure~\ref{fig:ccl}. We can observe that \ccl can provide similar efficiency to NCCL, e.g., 89\% of NCCL on 2 GB data. Considering that the tensor length of gradients in data parallelism is around 2 GB,  \ccl can effectively replace NCCL for cross-node communication with negligible cost, thereby directly benefiting training recovery.

\begin{figure}[t]
	\centering\includegraphics[width=1.0\columnwidth]{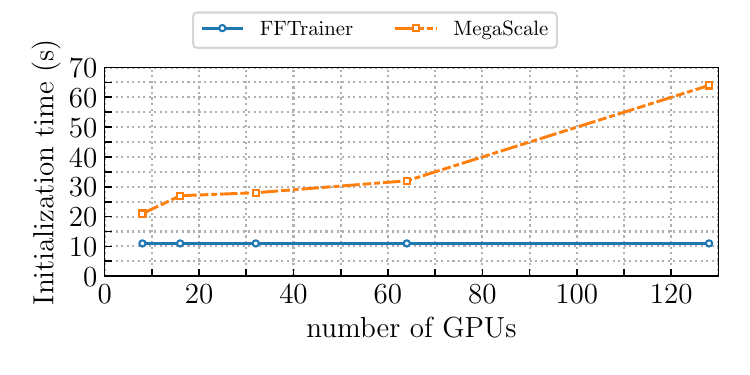}
	\caption{Network initialization overhead: \ccl vs. MegaScale for different scales.} 
	\label{fig:init}
\end{figure}

\para{\ccl outperforms MegaScale in network state recovery.}
We measure the time overhead of network state recovery at different scales using various communication libraries, as shown in Figure~\ref{fig:init}. \ccl completes recovery steps, such as connection building and group initialization, faster than MegaScale, which requires extra time to convert model tensors into DTensors~\cite{veScale}. For example, in 128-GPU clusters, \ccl takes only 17\% of the time required by MegaScale. Additionally, since \sysname overlaps network state recovery with other steps like model loading, the current overhead is hidden without impacting failover.



\subsection{Scalability and Future-proofing}

\begin{figure}[t]
	\centering\includegraphics[width=1.0\columnwidth]{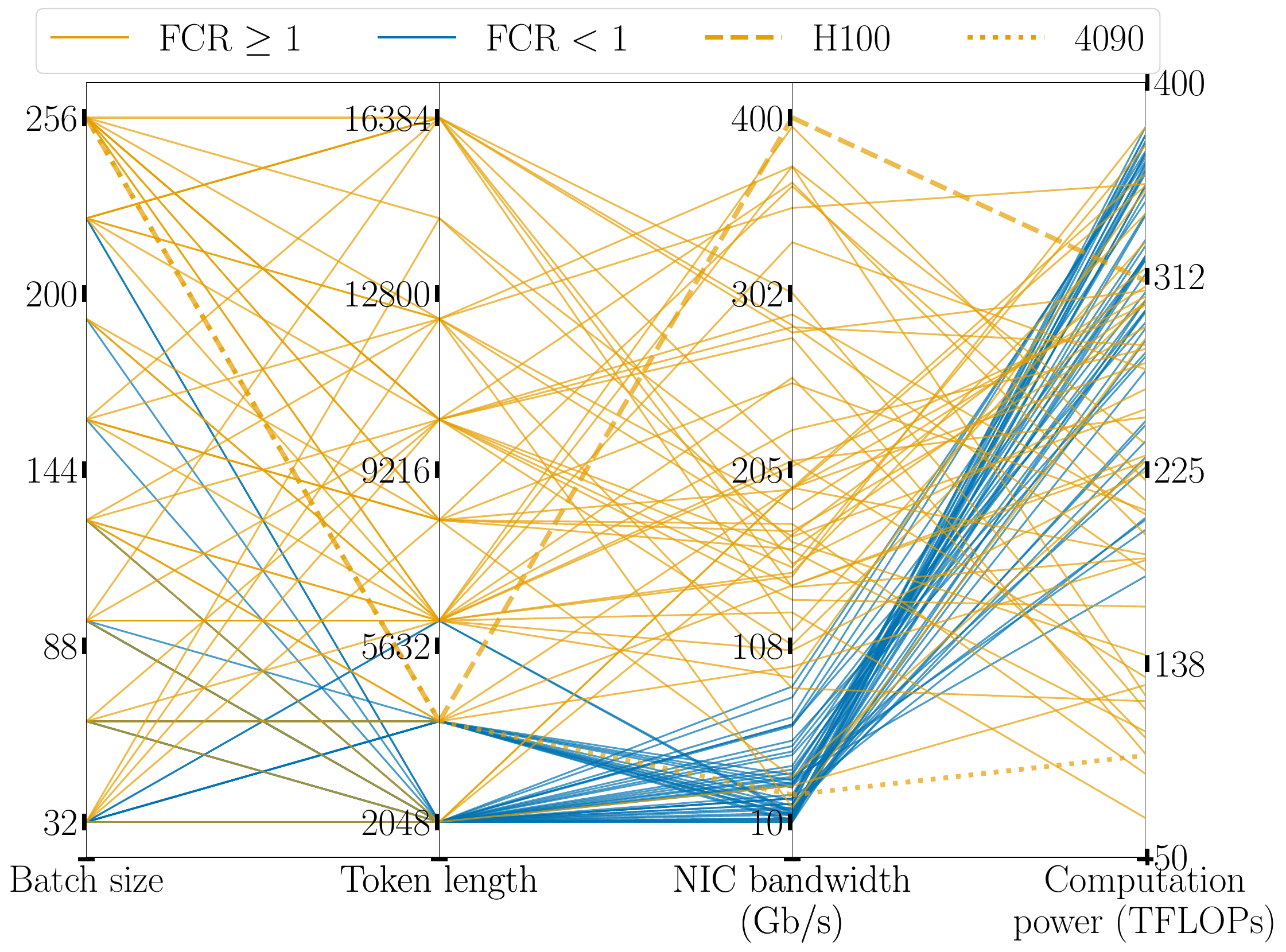}
	\caption{Factors influencing FCR. The orange lines represent factor combinations satisfying $\text{FCR} \geq 1$ (free {\ckpt}) while the blue lines represent $\text{FCR} < 1$ (paid {\ckpt}). The dashed lines show real cases of training LLaMA3 using 4090 and H100 GPUs in a batch size of 256.} 
	\label{fig:fcr}
\end{figure}

\para{\sysname can work on future models and devices.}
A critical assumption of \sysname is that we can complete all state management within a single iteration.  The key question is whether the assumption still holds on future models and computation platforms. 

We investigate the relationship between FCR (Section~\ref{sec:state}) and the training parameters (batch sizes and token length) and network/computation resources (per GPU network bandwidth and per-GPU FLOPS).  We sample a variety of combinations of these parameters, and use \emph{parallel coordinate plot}~\cite{parallelaxis} to illustrate them in Figure~\ref{fig:fcr}.  
Specifically, the orange line represents combinations that satisfy {\sysname}'s assumption ($\text{FCR} \geq 1$), while the blue line represents the cases where $\text{FCR} < 1$. The two thick lines indicate typical settings for the H100 and 4090 GPUs, respectively.  

We observe that the orange lines across many different settings, while blue lines are all clustered at small regions with small token lengths, very small bandwidth (less than 25Gbps), and fast GPUs (about $4\times$ higher than 4090s, or closer to H100).  These blue lines represent impractical settings where small models are trained on very fast GPUs interconnected by low-bandwidth networks. 

With the increasing scale of LLMs and computation power, the $\text{FCR} \geq 1$ assumption will become even stronger in the future:
(1) Maximum token lengths will increase, raising FCR;
(2) Larger GPU memory will allow for increased batch sizes, further boosting FCR; 
(3) While higher GPU FLOPS reduce FCR due to shorter network idle times, increasing network bandwidth will compensate for this. Even if GPUs become $82\times$ faster than H100, existing 400 Gbps NICs can still guarantee $\text{FCR} \geq 1$. 
In summary, future technology trends will only favor the FCR assumption of \sysname. 


\begin{figure}[t]
	\centering\includegraphics[width=1.0\columnwidth]{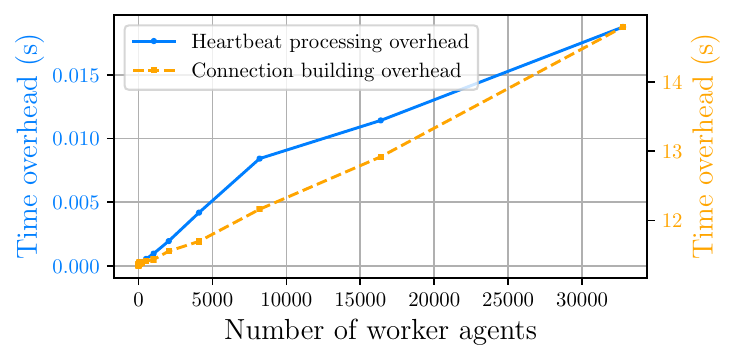}
	\caption{Time overhead of processing heartbeats and connection building by the state controller at different scales.} 
	\label{fig:scale}
\end{figure}

\para{\sysname scales to 30k+ GPUs. }
As we analyze in Section~\ref{sec:load}, the only scalability bottleneck is the centralized state controller.  To stress-test the controller and ensure it does not become an actual bottleneck, we create 32,768 workers on our testbed without using GPUs, each sending heartbeats and attempting to establish connections with the controller, and measure processing time on the controller. 

Figure~\ref{fig:scale} reveals the following conclusions:  
(1) It takes 11 seconds for a single worker to build connections (in order to register the RDMA buffer, etc.), and when we increase the number of workers to 32,768, it takes only 14 seconds in total (yellow line/axis); 
(2) Processing heartbeats from 32,768 workers only uses 19 ms of CPU time (blue line/axis); 
(3) Both times grow close to linearly with number of workers. 
Thus, the central controller is unlikely to become a bottleneck. 




\section{Conclusion and Future Work}
\label{sec:conclusion}
Despite substantial engineering to prevent failures, distributed LLM training systems are not failover-efficient; they rely on costly checkpointing that yields long recovery times. \sysname shifts focus from increasing MTBF to minimizing MTTR, cutting $98\%$ recovery time so frequent faults become tolerable. It delivers instant checkpointing by minimizing saved state (preventing rollbacks) and accelerates recovery via fast, asynchronous cross-node communication initialization and lightweight failure detection with \ccl. By routing and prioritizing state transfers on the training network, \sysname overlaps checkpointing with the transmission of training data, requiring no extra data network. \sysname remains effective on larger, faster future systems. As future work, we will explore checkpoint compression to further reduce overhead and mitigate FCR limitations.

\bibliographystyle{ACM-Reference-Format}
\bibliography{reference}

\end{document}